\documentclass[aps,prd,twocolumn,showpacs,preprintnumbers,superscriptaddress,nofootinbib,floatfix]{revtex4-1}
\usepackage{amsmath}
\usepackage{bm}
\usepackage{amsfonts}
\usepackage{graphicx}
\usepackage{epstopdf}
\usepackage{hyperref}
\usepackage{array}
\usepackage{booktabs}
\usepackage{soul}
\usepackage{color}
\enlargethispage{\baselineskip}

\hypersetup{
     colorlinks   = true,
     citecolor    = blue
}

\everymath{\displaystyle}

\newcommand{\MP}{M_{\rm PL}}
\newcommand{\Planck}{{\it Planck}~}

\renewcommand\({\left(}
\renewcommand\){\right)}
\renewcommand\[{\left[}
\renewcommand\]{\right]}
\newcommand{\be}{\begin{equation}}
\newcommand{\ee}{\end{equation}}
\newcommand{\bea}{\begin{eqnarray}}
\newcommand{\eea}{\end{eqnarray}}
\newcommand{\picLCDMH}{\includegraphics[width=0.8\linewidth]{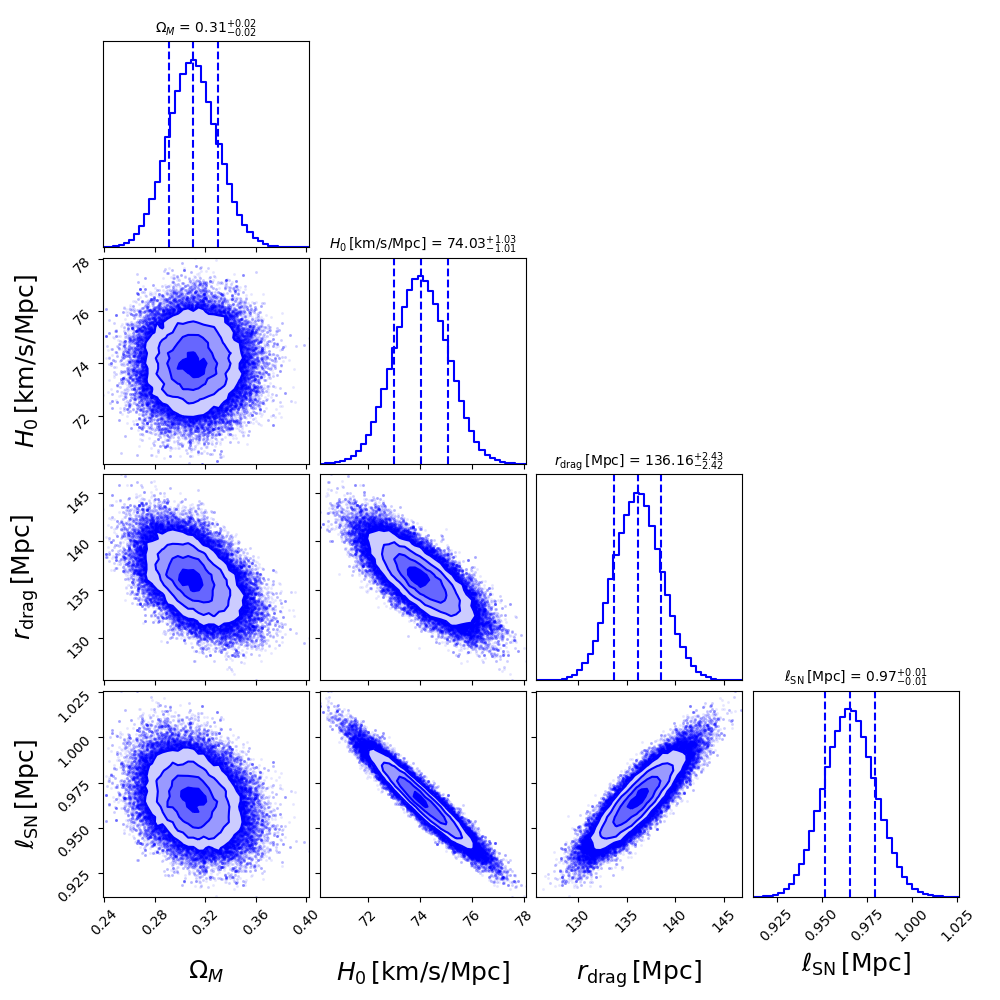}}
\newcommand{\picwCDMH}{\includegraphics[width=0.8\linewidth]{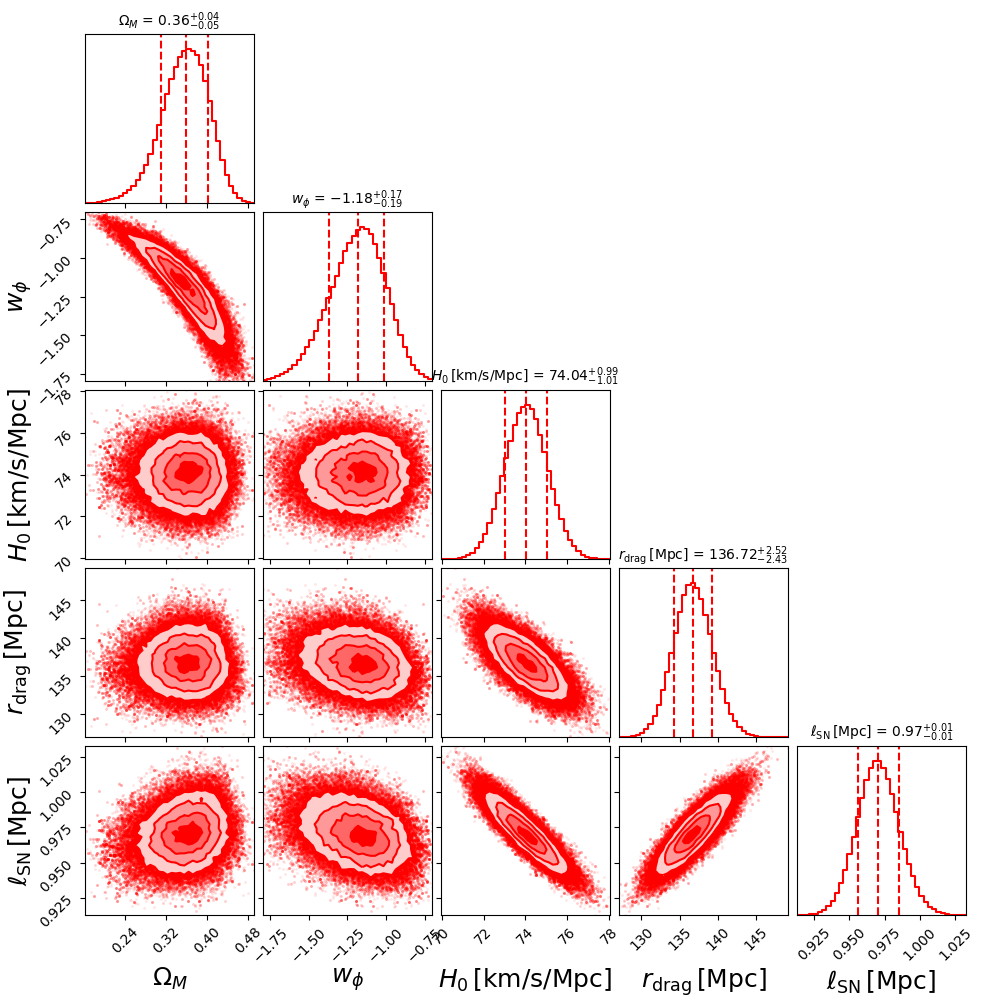}}
\newcommand{\piccCDMH}{\includegraphics[width=0.8\linewidth]{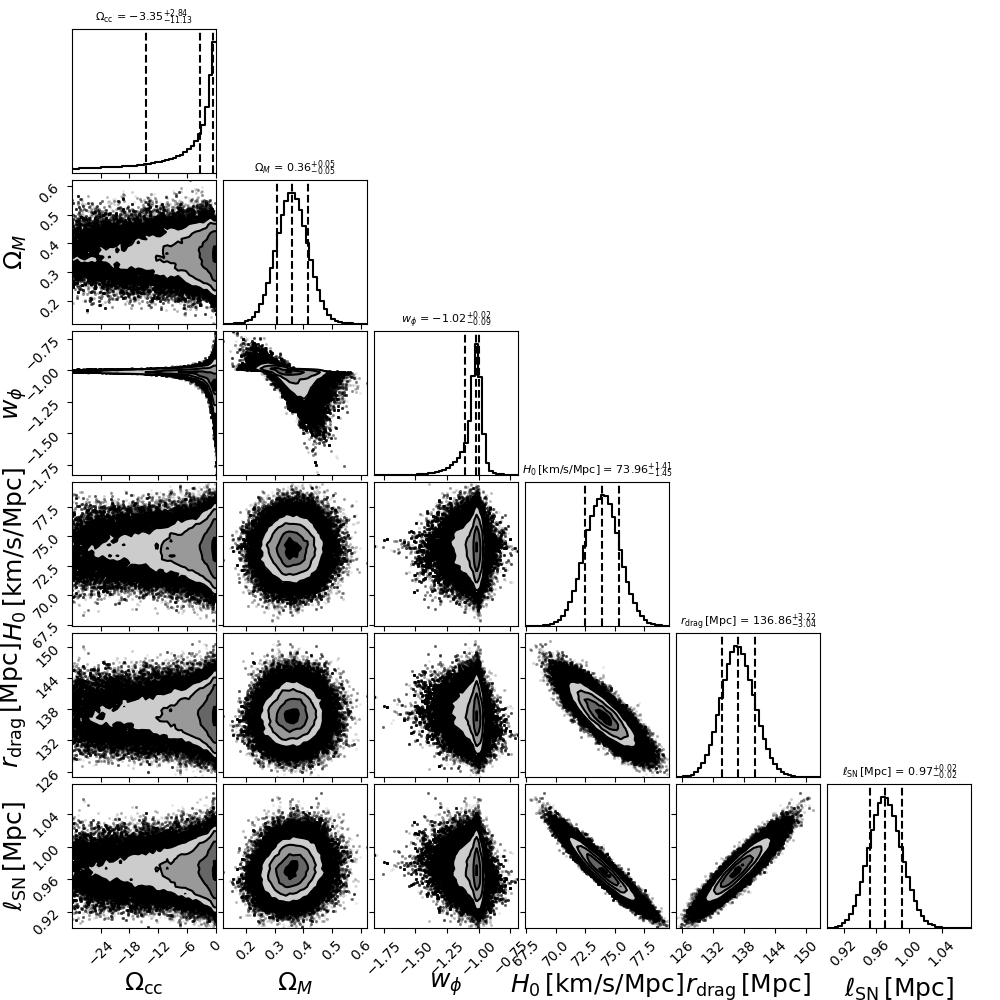}}
\newcommand{\picLCDMrd}{\includegraphics[width=0.8\linewidth]{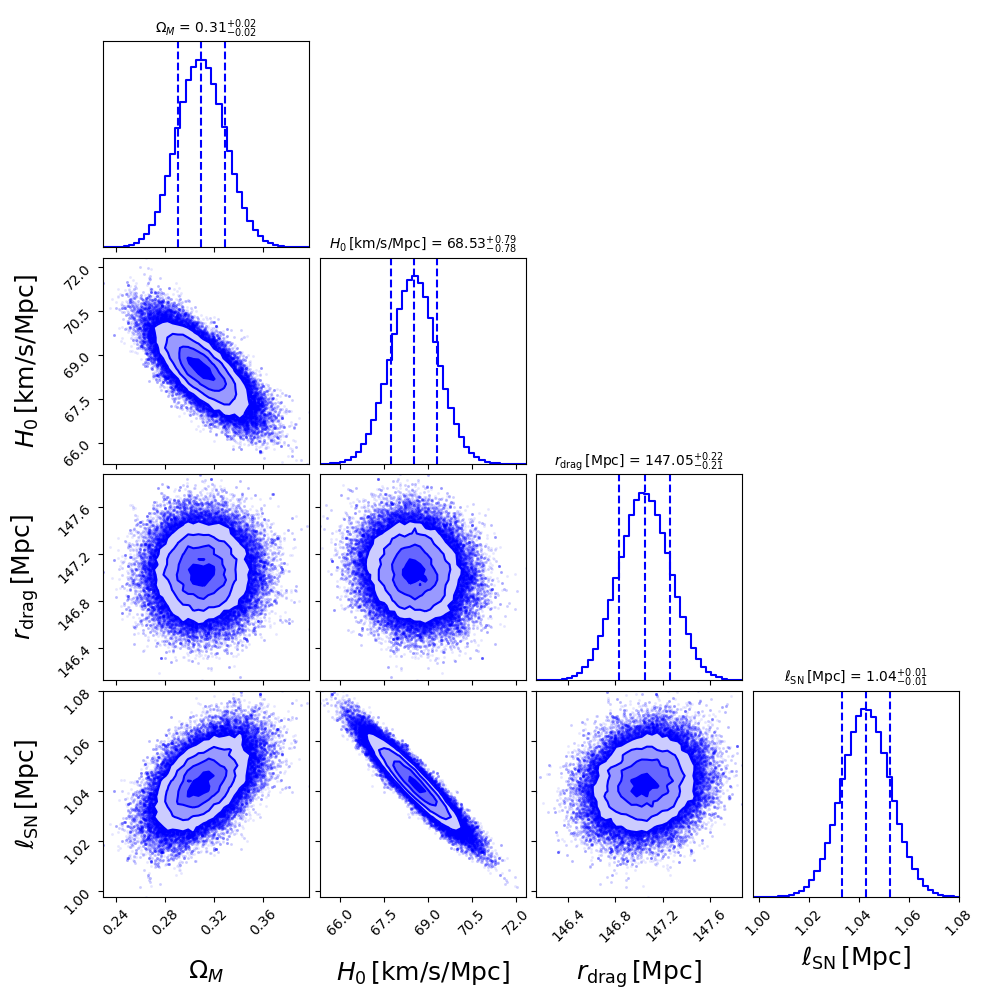}}
\newcommand{\picwCDMrd}{\includegraphics[width=0.8\linewidth]{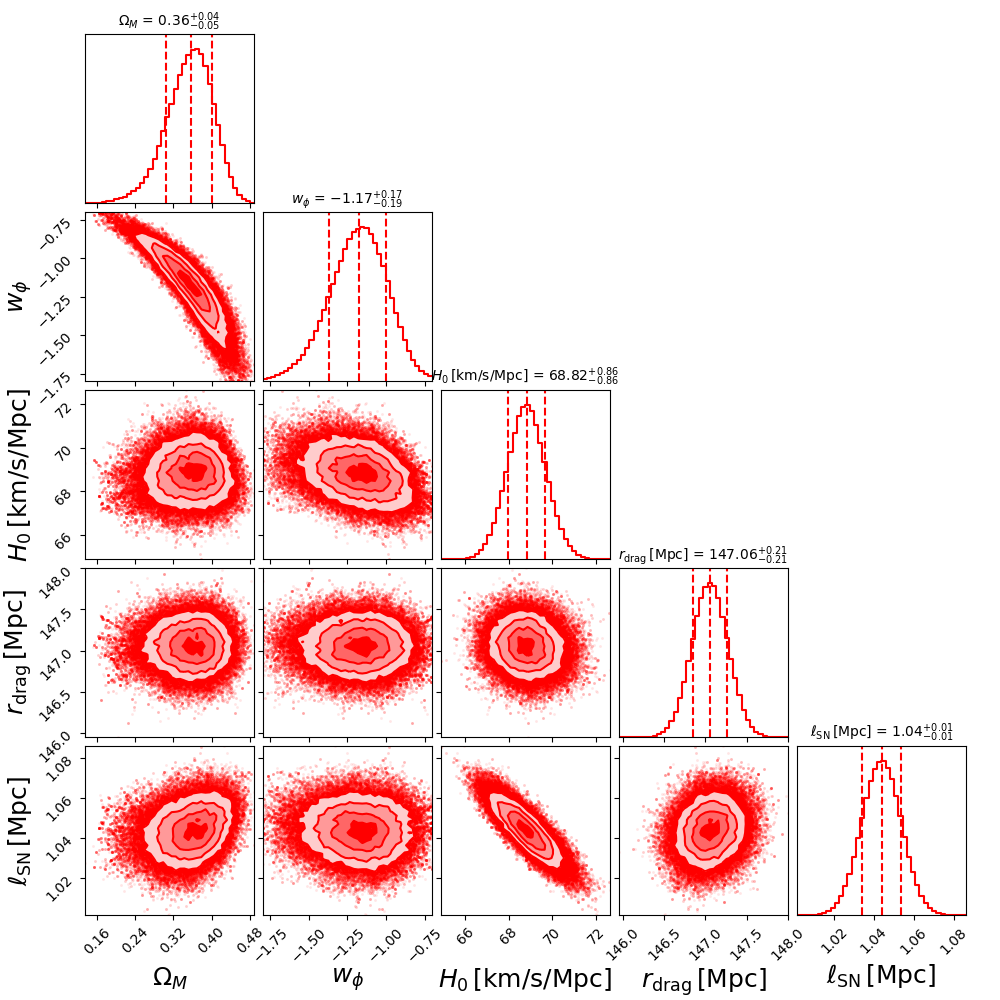}}
\newcommand{\piccCDMrd}{\includegraphics[width=0.8\linewidth]{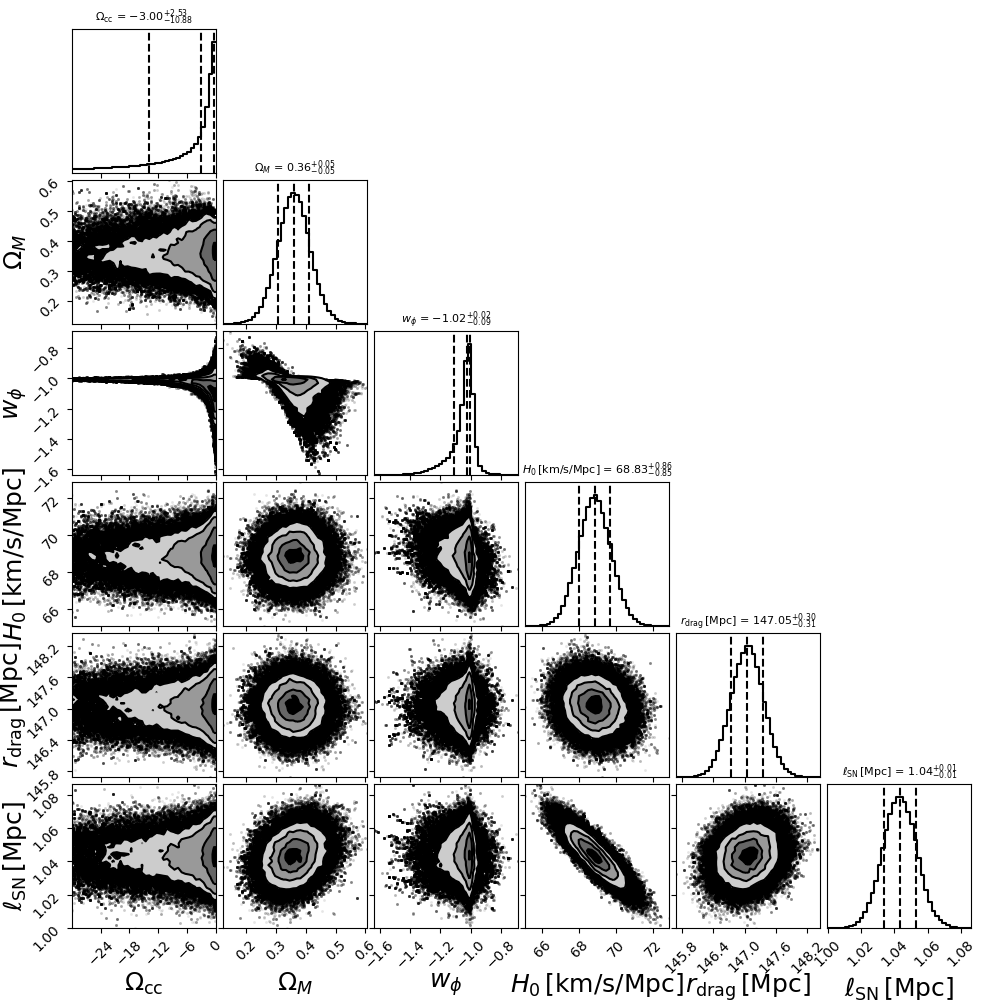}}
\newcolumntype{C}{>{\centering\arraybackslash}m{0.5\linewidth}}

\begin{document}
\title{Revisiting a negative cosmological constant from low-redshift data}

\newcommand{\FIRSTAFF}{\affiliation{Department of Physics and Astronomy, Uppsala University, L\"agerhyddsv\"agen 1, SE-751 20 Uppsala, Sweden}}
\newcommand{\SECONDAFF}{\affiliation{Nordita, KTH Royal Institute of Technology and Stockholm University, Roslagstullsbacken 23, SE-106 91 Stockholm, Sweden}}
\newcommand{\THIRDAFF}{\affiliation{The Oskar Klein Centre for Cosmoparticle Physics, Department of Physics, Stockholm University, \\AlbaNova Universitetscentrum, Roslagstullsbacken 21A, SE-106 91 Stockholm, Sweden}}
\newcommand{\FOURTHAFF}{\affiliation{Kavli Institute for Cosmology (KICC) and Institute of Astronomy,\\University of Cambridge, Madingley Road, Cambridge CB3 0HA, United Kingdom}}

\author{Luca Visinelli}\email[]{luca.visinelli@physics.uu.se} \FIRSTAFF \SECONDAFF
\author{Sunny Vagnozzi}\email[]{sunny.vagnozzi@fysik.su.se} \SECONDAFF \THIRDAFF \FOURTHAFF
\author{Ulf Danielsson}\email[]{ulf.danielsson@physics.uu.se}\FIRSTAFF

\date{\today}

\begin{abstract}

Persisting tensions between high-redshift and low-redshift cosmological observations suggest the dark energy sector of the Universe might be more complex than the positive cosmological constant of the $\Lambda$CDM model. Motivated by string theory, wherein symmetry considerations make consistent AdS backgrounds (\textit{i.e.} maximally symmetric spacetimes with a negative cosmological constant) ubiquitous, we explore a scenario where the dark energy sector consists of two components: a negative cosmological constant, with a dark energy component with equation of state $w_{\phi}$ on top. We test the consistency of the model against low-redshift Baryon Acoustic Oscillation and Type Ia Supernovae distance measurements, assessing two alternative choices of distance anchors: the sound horizon at baryon drag determined by the \textit{Planck} collaboration, and the Hubble constant determined by the SH0ES program. We find no evidence for a negative cosmological constant, and mild indications for an effective phantom dark energy component on top. A model comparison analysis reveals the $\Lambda$CDM model is favoured over our negative cosmological constant model. While our results are inconclusive, should low-redshift tensions persist with future data, it would be worth reconsidering and further refining our toy negative cosmological constant model by considering realistic string constructions.

\end{abstract}

\maketitle
\preprint{UUITP-28/19}

\section{Introduction}
\label{sec:intro}

While extremely successful at describing a wide variety of high- and low-redshift observations~\cite{Riess:1998cb,Perlmutter:1998np,Alam:2016hwk,Troxel:2017xyo,Aghanim:2018eyx}, the $\Lambda$CDM model has recently begun to display a number of small cracks~\cite{Freedman:2017yms,DiValentino:2017gzb}. One of the most tantalizing among these crevices is the so-called ``$H_0$ tension'', referring to the discrepancy between two independent estimates of the Hubble constant $H_0$. The first is a local estimate from the \textit{Hubble Space Telescope} (HST), based on a distance ladder approach using Cepheids variables in the hosts of Type Ia Supernovae (SNeIa), and yielding $H_0 = (74.03 \pm 1.42)\,{\rm km \,s^{-1}\,Mpc^{-1}}$~\cite{Riess:2019cxk}. The second is an indirect estimate based on Cosmic Microwave Background (CMB) temperature and polarization anisotropies measurements from the \textit{Planck} collaboration: assuming an underlying $\Lambda$CDM model, a value of $H_0 = (67.36 \pm 0.54)\,{\rm km \,s^{-1}\,Mpc^{-1}}$~\cite{Aghanim:2018eyx} is obtained.~\footnote{Besides this cosmological tension, the $\Lambda$CDM model also suffers from a number of astrophysical shortcomings at small (galactic or subgalactic scales), mostly referring to mismatches between simulations of cold dark matter and observations. Such issues include the core-cusp problem~\cite{deBlok:2009sp}, the missing satellites problem~\cite{Klypin:1999uc} (although see for example~\cite{Simon:2007dq,Kim:2017iwr}), and the too-big-to-fail problem~\cite{BoylanKolchin:2011de}. See for instance~\cite{Weinberg:2013aya,DelPopolo:2016emo,Bullock:2017xww} for recent reviews on the subject. It has been speculated that such issues might signal the need to move away from the collisionless cold dark matter paradigm, and in particular that allowing for interactions between dark matter particles, or between dark matter and baryons, could alleviate these issues (see e.g.~\cite{Spergel:1999mh,Strigari:2007ma,Aarssen:2012fx,Tulin:2013teo,Foot:2014uba,
Foot:2014osa,Kaplinghat:2015aga,Foot:2016wvj,Binder:2016pnr,Tang:2016mot,
Bringmann:2016ilk,Du:2016zcv,Tulin:2017ara,Deng:2018jjz,Robles:2018fur} for an incomplete list of proposed models).}

The statistical significance of this discrepancy, currently at a level $\gtrsim 4\sigma$, has steadily grown over recent year, owing also improvements in the distance ladder approach~\cite{Riess:2009aaa, Riess:2011aaa,Riess:2016jrr,Riess:2018uxu}. Additional (re)analyses of the local distance ladder~\cite{Efstathiou:2013via,Cardona:2016ems,Zhang:2017aqn,Feeney:2017sgx,Dhawan:2017ywl,Follin:2017ljs,Gomez-Valent:2018hwc,Burns:2018ggj,Collett:2019hrr,Camarena:2019moy,Freedman:2019abc}, as well as independent estimates of $H_0$ from strong-lensing time delays~\cite{Bonvin:2016crt,Birrer:2018vtm,Wong:2019kwg} have independently supported the conclusion that local measurements favour a higher value of $H_0$~\cite{Riess:2019cxk} (see, however, the study in~\cite{Lombriser:2019ahl}). Given that the \textit{Planck} estimate of $H_0$ relies on the assumption of an underlying $\Lambda$CDM model, it is possible that the $H_0$ tension might be hinting towards new physics, possibly involving non-standard properties of dark matter or dark energy~\cite{Bernal:2016gxb, DiValentino:2016hlg, Mortsell:2018mfj, Guo:2018ans}. The task is not easy to accomplish, since the simplest single-field quintessence models cannot solve the $H_0$ tension~\cite{Vagnozzi:2018jhn,DiValentino:2019dzu}, even when a late-time transition from a matter-like equation of state (EoS) is considered~\cite{DiValentino:2019exe}. For an incomplete list of works in this direction, see Refs.~\cite{Barreira:2014ija,Murgia:2016ccp,Qing-Guo:2016ykt,Tram:2016rcw,Ko:2016uft,Karwal:2016vyq,Kumar:2016zpg,Xia:2016vnp,
Chacko:2016kgg,Zhao:2017cud,Kumar:2017dnp,Agrawal:2017rvu,Benetti:2017gvm,
Feng:2017nss,Zhao:2017urm,DiValentino:2017zyq,Gariazzo:2017pzb,Dirian:2017pwp,
DiValentino:2017iww,Sola:2017znb,Feng:2017mfs,Renk:2017rzu,Yang:2017alx,Buen-Abad:2017gxg,Yang:2017ccc,Raveri:2017jto,DiValentino:2017rcr,DiValentino:2017oaw,
Khosravi:2017hfi,Santos:2017alg,Peirone:2017vcq,Benetti:2017juy,Feng:2017usu,Belgacem:2017cqo,
Vagnozzi:2018jhn,Nunes:2018xbm,Poulin:2018zxs,Kumar:2018yhh,Yang:2018ubt,
Yang:2018euj,Yang:2018xlt,Banihashemi:2018oxo,DEramo:2018vss,deMartino:2018cjh,Graef:2018fzu,Yang:2018uae,
Benevento:2018xcu,Lin:2018nxe,Yang:2018qmz,Yang:2018xah,Banihashemi:2018has,
Aylor:2018drw,Chiang:2018xpn,Poulin:2018cxd,Carneiro:2018xwq,Lazkoz:2019ivd,Kreisch:2019yzn,
Pandey:2019plg,Kumar:2019wfs,Vattis:2019efj,Pan:2019jqh,Colgain:2019pck,
Agrawal:2019lmo,Alexander:2019rsc,Yang:2019jwn,Adhikari:2019fvb,Keeley:2019esp,
Peirone:2019yjs,Lin:2019qug,Yang:2019qza,Cerdonio:2019oqr,Agrawal:2019dlm,Li:2019yem,
Gelmini:2019deq,DiValentino:2019exe,Yang:2019uzo,Archidiacono:2019wdp,Desmond:2019ygn,
Yang:2019nhz,Nesseris:2019fwr,Baldes:2019tkl,Pan:2019gop,Vagnozzi:2019ezj,Cai:2019bdh,
Schoneberg:2019wmt,Pan:2019hac}.

Measurements of temperature and polarization anisotropies in the CMB exquisitely constrain $\theta_s$, the angular scale of the sound horizon at last-scattering:
\begin{eqnarray}
\theta_s = \frac{r_s(z_{\rm drag})}{D_A(z_{\star})}\,,
\label{eq:angle}
\end{eqnarray}
where $z_{\star} \approx 1100$ is the redshift of last-scattering, $z_{\rm drag} \approx z_{\star}$ is the redshift of the drag epoch (when baryons are released from the Compton drag of photons), $r_s(z)$ is the comoving sound horizon at redshift $z$, and $D_A(z)$ is the angular diameter distance to redshift $z$. Attempts to address the $H_0$ tension by introducing new physics generally modify either or both $r_s$ and $D_A$, in such a way that an increase in $H_0$ is required in order to keep $\theta_s$ fixed. One of the simplest possibilities is to modify the dark energy (DE) sector in order to lower the expansion rate at late times, while leaving the early-time expansion rate unaltered. Such a change will keep $r_s(z_{\rm drag})$ unchanged, but will raise $D_A(z_{\star})$. In order to keep $\theta_s$ fixed, a higher value of $H_0$ will be inferred.

One of the simplest modifications to the DE sector that could be considered consists in replacing the positive cosmological constant (whose equation of state is $w_{\rm cc} = -1$) with a quintessence field with a time-varying EoS. A single minimally coupled quintessence field $\phi$ with standard kinetic term possesses a positive energy density with an equation of state $w_\phi(z) \geq -1$; it is easy to show that, in this case, the expansion rate actually increases at late times, going in the opposite direction of what is required to address the $H_0$ tension~\cite{Vagnozzi:2018jhn}. In fact, insisting on a positive energy density for the DE component brings one to consider \textit{phantom} DE, \textit{i.e.} a DE component with equation of state $w_\phi < -1$. Phantom DE components, which generically predict the Universe to end in a Big Rip~\cite{Caldwell:2003vq} (although this fate can be avoided in certain classes of modified gravity theories~\cite{Nojiri:2005sx,Zhang:2009xj,Bamba:2009uf,Astashenok:2012tv,Mortsell:2017fog}), are however generically extremely challenging to construct, due to the violation of the strong energy condition. Nonetheless, attempts to realize effectively stable phantom DE components (for instance within modified gravity or brane-world models) exist, see e.g. Refs.~\cite{Damour:1990tw,Sahni:2002dx,Torres:2002pe,Chung:2002xj,Faraoni:2003jh,Carroll:2004hc,Das:2005yj,Jhingan:2008ym,Nojiri:2013ru,Ludwick:2015dba,Cognola:2016gjy,Sebastiani:2016ras,Ludwick:2017tox,Barenboim:2017sjk,Casalino:2018tcd}.

The situation changes if one chooses not to restrict to the DE energy density being positive. Recent work has in fact shown that a number of persisting low-redshift tensions (including the $H_0$ tension) might be addressed if one allows for an evolving DE component whose energy density can assume negative values (see e.g. Refs.~\cite{Mortsell:2018mfj,Poulin:2018zxs,Capozziello:2018jya,Wang:2018fng,Dutta:2018vmq}). In particular, Ref.~\cite{Dutta:2018vmq} considered a very interesting case, where the DE sector consists of a slowly-rolling quintessence field (whose energy density is positive) on top of a negative cosmological constant. Such a scenario is extremely interesting from a string theory perspective. In fact, constructing meta-stable de Sitter (dS) vacua (\textit{i.e.} with a positive cosmological constant) has notoriously been a daunting task in string theory (see e.g. Refs.~\cite{Maldacena:2000mw,Silverstein:2007ac,Danielsson:2009ff,Wrase:2010ew,
Danielsson:2010bc,Danielsson:2011au,Chen:2011ac,
Danielsson:2012et,Dasgupta:2014pma,Cicoli:2018kdo}). These difficulties have led to the suggestion that string theory might not have any dS vacua at all~\cite{Vafa:2005ui,Danielsson:2018ztv,Obied:2018sgi,Ooguri:2018wrx,Palti:2019pca}, an observation which, if correct, would have interesting consequences for cosmology~\cite{Agrawal:2018own,Andriot:2018wzk,Achucarro:2018vey,Garg:2018reu,Kehagias:2018uem,Heisenberg:2018yae,Kinney:2018nny,Akrami:2018ylq,Murayama:2018lie,Kinney:2018kew} (see however also the important work of~\cite{Kachru:2003aw}, which appears to provide a possible counterexample to these swampland conjectures). In contrast, a negative cosmological constant (providing an AdS background, \textit{i.e.} a maximally symmetric spacetime with a negative cosmological constant) is very natural from symmetry considerations in string theory, as can be argued using the AdS/CFT correspondence~\cite{Maldacena:1997re}. Contrary to the case of dS, there does seem to exist a large number of consistent AdS backgrounds that can be obtained from string theory (it has even been suggested that AdS vacua can lead to an accelerating Universe~\cite{Hartle:2012qb}).

Another rather generic prediction of string theory is the plethora of light bosons known as the ``axiverse''. These correspond to moduli determining the size and shape of the extra dimensions, and could give rise to important observable consequences that might alter the evolution of the Universe between recombination and today~\cite{Svrcek:2006yi,Arvanitaki:2009fg,Arvanitaki:2010sy,Cicoli:2012sz,
Conlon:2013isa,Cicoli:2014bfa,Marsh:2015xka,Visinelli:2017imh,Poulin:2018dzj,
Kitajima:2018zco,Visinelli:2018utg,Odintsov:2019mlf,Odintsov:2019evb,
Ramberg:2019dgi,Reynolds:2019uqt}. It is therefore interesting to try matching cosmological data using quintessence (possibly with an effective phantom equation of state, given that one is anyhow not dealing with a single quintessence field) in combination with a negative cosmological constant, both of which are natural from the point of view of string theory. This is an important motivation for our paper.

At any rate, determining the sign of a possible cosmological constant is crucial in order to understand the structure and fabric of space-time itself. In this work, we revisit the possibility that a negative cosmological constant might be allowed by current cosmological data, and more generally explore whether the $H_0$ tension might be relaxed within such a scenario. We first consider the standard $\Lambda$CDM model (featuring a positive cosmological constant with equation of state $w=-1$). We then move on to one of the simplest extension of the $\Lambda$CDM model, \textit{i.e.} the $w$CDM model, wherein the DE EoS $w$ is treated as a free parameter. Finally, we consider a scenario in which the present accelerated expansion rate is driven by a combination of a cosmological constant whose density parameter $\Omega_{cc}$ is strictly negative, on top of which we allow for a second DE component with EoS $w_{\phi}$ (which can take on values $w_{\phi}<-1$) and a positive density parameter $\Omega_{\phi}$. While the case for a negative cosmological constant is clearly inspired by a string scenario, the present accelerated expansion rate requires that $\Omega_{\rm cc} + \Omega_\phi > 0$. We refer to this last model as $c$CDM. Notice that the requirement $\Omega_{\rm cc} + \Omega_\phi > 0$, necessary in order for the Universe to accelerate, prevents us from considering a model where the dark energy sector consists solely of a negative cosmological constant, without the quintessence field on top.

In this work, we perform a Markov Chain Monte Carlo (MCMC) analysis to compare the three models (in order of increasing complexity $\Lambda$CDM, $w$CDM, and $c$CDM) in light of recent low-redshift ($z \lesssim 2.5$) distance measurements from a selection of Baryon Acoustic Oscillation (BAO) surveys~\cite{Beutler:2011hx,Ross:2014qpa,Alam:2016hwk,Bourboux:2017cbm,
Bautista:2017zgn,Ata:2017dya} and Type Ia Supernovae (SNeIa) from the Pantheon catalogue~\cite{Scolnic:2017caz}. The interpretation of BAO measurements requires ``anchoring'' them to either the cosmic distance ladder through independent measurements of $H_0$, or to the inverse distance ladder through independent measurements of $r_{\rm drag}$, the sound horizon at baryon drag~\cite{Bernal:2016gxb}. To show the impact of the $H_0$ tension, we choose two different anchors to interpret our BAO data: we first use the local distance ladder measurement of $H_0$ from the ``Supernovae and $H_0$ for the Dark Energy Equation of State'' (SH0ES) program~\cite{Riess:2019cxk}, before comparing our results to those obtained anchoring to the most recent measurement of $r_{\rm drag}$ from the \textit{Planck} collaboration~\cite{Aghanim:2018eyx}. More details on the datasets chosen and the choice of anchoring are given in Sec.~\ref{sec:datasets}.

Our treatment of the effective DE component on top of the negative cosmological constant is purely phenomenological, we remain agnostic as to the underlying Lagrangian for such a component (see e.g.~\cite{Grande:2006nn,Grande:2006qi} for early works discussing a fundamental origin for this type of composite dark energy). While we envisage such a component being a quintessence field slowly rolling along a potential on top of the negative cosmological constant, we allow the effective equation of state $w_{\phi}$ to enter the phantom regime, where $w_{\phi}<-1$. The rationale is that string constructions generically predict a large number of light bosons on top of the stable AdS vacua: therefore, one can in general be faced with a multi-field quintessence scenario, whose \textit{effective} behaviour might be phantom (see e.g. Refs.~\cite{Guo:2004fq,Hu:2004kh,Cai:2009zp,Saridakis:2009jq,Saridakis:2016ahq}).~\footnote{Other possibilities for obtaining an effective phantom component from an underlying scalar field model involve considering modifications to gravity~\cite{Amendola:2007nt,Saridakis:2009uk,Wu:2010av,Elizalde:2011ds,
Saridakis:2012jy,Pan:2015eta,Dhawan:2017leu,Elizalde:2017mrn,
Odintsov:2017hbk,Dutta:2017fjw}, couplings between dark energy and dark matter~\cite{Amendola:1999er,Barrow:2006hia,He:2008tn,Valiviita:2008iv,
Gavela:2009cy,Martinelli:2010rt,Pan:2012ki,Tamanini:2015iia,Nunes:2016dlj,
Pan:2016ngu,Sharov:2017iue,Pan:2017ent,Yang:2018qec,Paliathanasis:2019hbi,Yang:2019bpr}, particle creation mechanisms~\cite{Lima:2008qy,Lima:2009ic,Pan:2014lua,Nunes:2015rea,deHaro:2015hdp,
Pan:2016jli,Nunes:2016aup,Pan:2016bug,Paliathanasis:2016dhu,Pan:2018ibu}, or invoking scalar fields non-minimally coupled to gravity or with a kinetic term which is non-canonical~\cite{Vikman:2004dc,Creminelli:2008wc,Deffayet:2010qz,
Akrami:2017cir,Paliathanasis:2018vru}.} In the $c$CDM model, the quintessence field (whose effective EoS $w_{\phi}$ we allow to be phantom, again from a purely effective perspective) is treated as an additional contribution to the total energy density along with the (possibly negative) cosmological constant and the usual components of matter and radiation. Our results are interesting from a model-building perspective, as we find that from the low-redshift data side a negative cosmological constant is a valid alternative to a phantom dark energy component.

The rest of the paper is then organized as follows. In Sec.~\ref{sec:datasets} we describe the datasets we use. In Sec.~\ref{sec:models}, we describe in detail the models considered along with their parameter spaces. The MCMC analysis performed in order to constrain the models, and the obtained results, are described in Sec.~\ref{sec:results}. We provide concluding remarks in Sec.~\ref{sec:summary}.

\section{Overview of the datasets used}
\label{sec:datasets}

The evolution of the Hubble expansion rate on the redshift $z$, $H(z)$, is usually factored out as $H(z) = H_0\,E(z)$, where $H_0$ is the Hubble constant and the normalized expansion rate $E(z)$ is a dimensionless function describing the evolution of the expansion rate with the redshift. Let us introduce the angular diameter distance to redshift $z$, $D_A(z)$, given by the following:
\be
	D_A(z) \equiv \frac{1}{1+z}\frac{c}{H_0}\int_0^z\frac{dz'}{E(z')}\,.
	\label{eq:da}
\ee
The angular diameter distance in Eq.~(\ref{eq:da}) is related to the comoving angular diameter distance to redshift $z$, $D_M(z)$, by $D_A(z) = D_M(z)/(1+z)$. Let us further define the sound horizon at redshift $z$, $r_s(z)$, as the distance travelled by an acoustic wave in the baryon-photon plasma from a very early time, given by:
\be
	r_s(z) = \int_z^{\infty} \frac{c_s(z')}{H(z')}dz'\,,
\ee
where $c_s(z)$ is the sound speed of the baryon-photon plasma at redshift $z$.

We consider two classes of observational datasets. A first class consists of distance measurements from Baryon Baryon Acoustic Oscillations (BAO) surveys as well as from Type Ia Supernovae (SNeIa). A second class consists of two different types of anchors used to interpret the BAO measurements. Specifically, we consider two different anchors: direct measurements of the Hubble constant based on the local Cepheids-Supernovae distance ladder, and CMB measurements of the comoving sound horizon at baryon drag, further defined in Eq.~\eqref{eq:definerdrag} below. We describe these four datasets in more detail below

\begin{itemize}
\item{\bf BAO.} In the early Universe, the interplay between gravity and radiation pressure sets up acoustic oscillations which produce a sharp feature in the two-point correlation function of luminous matter at a scale equal to the comoving size of the sound horizon at the drag epoch~\cite{Peebles:1970ag,Sunyaev:1970eu,Sunyaev:1970er,Eisenstein:2005su,Aubourg:2014yra}:
\be
	r_{\rm drag} \equiv r_s(z_{\rm drag})\,,
	\label{eq:definerdrag}
\ee
where the drag epoch is the time when the baryons are released from the Compton drag of the photons, and occurs at a redshift $z_{\rm drag}$. Measurements of the BAO feature, first reported in Refs.~\cite{Eisenstein:2005su,Cole:2005sx}, are usually performed at an effective redshift $z_{\rm eff}$. These measurements can in principle independently constrain the angular diameter distance $D_A(z_{\rm eff})$ in units of $r_{\rm drag}$ (for modes in the transverse direction with respect to the line of sight) and $H(z_{\rm eff})r_{\rm drag}$ (for modes along the line of sight), whereas isotropic BAO measurements constrain a combination of these quantities known as the volume distance $D_V(z_{\rm eff})$. Here, we consider various types of BAO measurements, which constrain the following quantities:
\bea
	\frac{D_M(z_{\rm eff})}{r_{\rm drag}} &\equiv& \frac{c}{H_0r_{\rm drag}}\int_0^{z_{\rm eff}}\frac{dz'}{E(z')}\,,\label{eq:deltaL} \\
	\frac{D_H(z_{\rm eff})}{r_{\rm drag}} &\equiv& \frac{c}{H_0r_{\rm drag}}\,\frac{1}{E(z_{\rm eff})}\,, \label{eq:deltaH}\\
	\frac{D_V(z_{\rm eff})}{r_{\rm drag}} &\equiv& \frac{c}{H_0\,r_{\rm drag}}\,\sqrt[3]{\frac{z_{\rm eff}}{E(z_{\rm eff})} \,\(\int_0^{z_{\rm eff}}\frac{dz'}{E(z')}\)^2}\,, \label{eq:deltaV}
\eea
See Appendix B of Ref.~\cite{Yang:2019vni} for a more detailed discussion of BAO measurements.

In our analysis, we consider anisotropic BAO measurements from the Baryon Oscillation Spectroscopic Survey (BOSS) collaboration~\cite{Alam:2016hwk} Data Release 12 (DR12) at the effective redshifts $z_{\rm eff} = 0.38\,,0.51\,,0.61$ and from Lyman-$\alpha$ forest samples at $z_{\rm eff} = 2.40$~\cite{Bourboux:2017cbm,Bautista:2017zgn}. We also include isotropic BAO measurements from the Six-degree Field Galaxy Survey (6dFGS) at $z_{\rm eff} = 0.106$~\cite{Beutler:2011hx}, from the Sloan Digital Sky Survey Data (SDSS) Main Galaxy Sample (MGS) release at $z_{\rm eff} = 0.15$~\cite{Ross:2014qpa}, and from the quasar sample of the extended Baryon Oscillation Spectroscopic Survey (eBOSS) at $z_{\rm eff} = 1.52$~\cite{Ata:2017dya}. For completeness, we have collected the BAO measurements used in this work in Table~\ref{table:measureBAO}. We construct the likelihood for the BAO data, $\mathcal{L}_{\rm BAO}$, using the data described and the correlation matrices provided by the collaborations.

\begin{table*}
	\begin{center}
	{\renewcommand{\arraystretch}{1.4}
	\begin{tabular}{|l|l|l|l|}
	\hline
		Dataset~\hspace{2cm} & $z~\hspace{0.5cm}$ & Measurement~\hspace{0.5cm} & Reference\\
	\hline
		BOSS DR12 & $0.38$ & $H\,\(r_{\rm drag}/r_{\rm fid}\) = \(81.2 \pm 2.2 \pm 1.0\){\rm \,km/(s\,Mpc)}$ & \cite{Alam:2016hwk}\\
		BOSS DR12 & $0.51$ & $H\,\(r_{\rm drag}/r_{\rm fid}\)  = \(90.9 \pm 2.1 \pm 1.1\){\rm \,km/(s\,Mpc)}$ & \cite{Alam:2016hwk}\\
		BOSS DR12 & $ 0.61$ & $H\,\(r_{\rm drag}/r_{\rm fid}\) = \(99.0 \pm 2.2 \pm 1.2\){\rm \,km/(s\,Mpc)}$ & \cite{Alam:2016hwk}\\
		BOSS DR12 & $0.38$ & $D_M/r_{\rm drag} = \(1512 \pm 22 \pm 11\){\rm \,Mpc}/r_{\rm fid}$ & \cite{Alam:2016hwk}\\
		BOSS DR12 & $0.51$ & $D_M/r_{\rm drag} = \(1975 \pm 27 \pm 14\){\rm \,Mpc}/r_{\rm fid}$ & \cite{Alam:2016hwk}\\
		BOSS DR12 & $ 0.61$ & $D_M/r_{\rm drag} = \(2307 \pm 33 \pm 17\){\rm \,Mpc}/r_{\rm fid}$ & \cite{Alam:2016hwk}\\
		BOSS DR12+Ly$\alpha$ & $2.40$ & $D_M/r_{\rm drag} = 36.6 \pm 1.2$ & \cite{Bourboux:2017cbm}\\
		BOSS DR12+Ly$\alpha$ & $2.40$ & $D_H/r_{\rm drag} = 8.94 \pm 0.22$ & \cite{Bourboux:2017cbm}\\
		6dFGS & $0.106$ & $D_V/r_{\rm drag} = 2.98 \pm 0.13$ & \cite{Beutler:2011hx}\\
		SDSS-MGS & $0.15$ & $D_V/r_{\rm drag} = 4.47 \pm 0.17$ & \cite{Ross:2014qpa}\\
		eBOSS quasars & $1.52$ & $D_V/r_{\rm drag} = 26.1 \pm 1.1$ & \cite{Ata:2017dya}\\
	\hline
	\end{tabular}}
	\caption{BAO scale measurements used in this work. For BOSS DR12, we set the fiducial scale $r_{\rm fid} = 147.78\,$Mpc~\cite{Alam:2016hwk}.}
	\label{table:measureBAO}
	\normalsize
	\end{center}
\end{table*}
\item{ \bf SNeIa.} SNeIa catalogues report the corrected apparent magnitude of the $i$-th supernova, which is given by the following:
\be
	\mu^i = M + 5\,{\rm Log}_{10}\(D_A^i(z)/{\rm Mpc}\) + 25\,,
\ee
where $M$ is the absolute magnitude of the $i$th supernova, which fixes the scale for the global fit. For practical purposes we rewrite the relation as
\be
	\mu^i = 5.7 + 5\,{\rm Log}_{10}\(\frac{c(1+z)}{H_0 \ell_{\rm SN}}\int_0^z\frac{dz'}{E(z')}\),
\ee
where the absolute magnitude has been reabsorbed into a parameter $\ell_{\rm SN} = 10^{-(M +19)/5}\,$Mpc which we fit to data. We construct the likelihood 
\be
	\mathcal{L}_{\rm SN} = \exp\[-\frac{1}{2}\(\mu - \mu{\rm obs}\)^T C_{\rm SN}^{-1}\(\mu - \mu_{\rm obs}\)\],
\ee
where the vector of apparent magnitude measurements $\mu_{\rm obs}$ and the corresponding covariance matrix $C_{\rm SN}^{-1}$ are taken from the Pantheon SNeIA dataset~\cite{Scolnic:2017caz}. This catalogue provides luminosity distances $D_A(z)$ for 1048 SNeIa within the range $0.01 < z \leq 2.3$ of Type 1a supernovae. Although the Pantheon analysis is marginally model-dependent, as biases in light-curve parameters are corrected assuming a $\Lambda$CDM cosmology, independent analyses run using the JLA SN dataset~\cite{Betoule:2014frx} find that the model-dependence of the light-curve parameters is weak.
\item {\bf Anchors} The cosmic distance ladder technique relies on measuring distances of extra-galactic objects, at distances beyond $\sim 100\,$Mpc, to map the Hubble flow. The expansion history of the universe is mapped through Type 1a supernovae (SNeIa) datasets~\cite{Riess:2006fw, 2009ApJ...700..331H, 2011ApJS..192....1C, 2012ApJ...746...85S, Betoule:2014frx, Sako:2014qmj, Jones:2017udy, Scolnic:2017caz} and BAO~\cite{Eisenstein:2005su, Cole:2005sx}. These distance scales need to be calibrated through anchors either at the high-redshift end (through $r_{\rm drag}$) or at the low-redshift end (through $H_0$)~\cite{Cuesta:2014asa}.

For instance, BAO measurements constrain the combination $H_0r_{\rm drag}$. In fact, the tension between measurements at low and high redshifts of $H_0$ can be eased by modifying the value of $r_{\rm drag}$, which is the standard ruler providing the BAO length scale. The sole measurements of either $H_0$ or $r_{\rm drag}$ are respectively interpreted as anchoring the cosmic distance ladder or the inverse cosmic distance ladder. In any case, in order to interpret the BAO measurements we have to anchor either $H_0$ or $r_{\rm drag}$ to an independent evaluation~\cite{Bernal:2016gxb}. Here we consider independent approaches, namely calibrating the cosmic distance ladder by means of the recent measurements of $H_0$ in Ref.~\cite{Riess:2019cxk}, or calibrating the inverse distance ladder by using the CMB measurement of $r_{\rm drag}$ in~\cite{Aghanim:2018eyx}. It is worth remarking that a lot (but not all) of the information contained in the CMB temperature and anisotropy spectra resides in the position and height of the first acoustic peak, which accurately constrains $\theta_s$ given in Eq.~(\ref{eq:angle}), and provides valuable information about the geometry and the content of the Universe. The position of the first peak is sensitive to early-time modifications through a change in $r_{\rm drag}$, as well as to late-time modifications through a change in $D_A(z_{\star})$.

Here, we consider two scenarios using two different anchors, given by the following:
\begin{enumerate}
\item {\bf Sound horizon at drag epoch (Scenario I).} The anchor at the CMB epoch is expressed by the comoving sound horizon at the end of the baryon drag epoch $r_{\rm drag}$. Measurements of CMB temperature and polarization anisotropies by the \Planck collaboration (\textit{PlanckTTTEEE}+\textit{lowP} dataset) give $r_{\rm drag} = \(147.05 \pm 0.30\)\,$Mpc at 68\% confidence level (C.L.). When anchoring data to the sound horizon $r_{\rm drag}$, we construct the likelihood function $\mathcal{L}_{\rm Anchor} = \mathcal{L}_{\rm drag}$ as a Gaussian in $r_{\rm drag}$ using this measurement.
\item {\bf Hubble constant $H_0$ (Scenario II).} When anchoring data to the present value of the Hubble rate, we use the latest result from the SH0ES program~\cite{Riess:2019cxk} which reports the local measurement $H_0 = \(74.03 \pm 1.42\)\,{\rm km \,s^{-1}\,Mpc^{-1}}$. This measurement is used to construct the likelihood function $\mathcal{L}_{\rm Anchor} = \mathcal{L}_{\rm Ceph}$ as a Gaussian in $H_0$.
\end{enumerate}

\end{itemize}

Since one of our main interests is to compare the results obtained using different anchors, we have not used other available datasets such as direct determinations of the Hubble rate $H(z)$ at redshifts $0.1 \lesssim z \lesssim 2$~\cite{Meng:2015loa, Moresco:2016mzx}, or ``compressed'' CMB likelihoods~\cite{Kosowsky:2002zt, Wang:2007mza, Dhawan:2017leu}. For ease of comparison to related works, we choose to restrict our use of low-redshift data solely to BAO and SNe data, which are the two most robust low-redshift datasets widely used by the community. We do not include other late-time measurements of the expansion history, such as cosmic chronometers or distance measurements from Gamma-ray bursts. However, it would be interesting to further include these datasets, which could perhaps improve our constraints on a possible negative cosmological constants, and we leave this exercise for future work.

\section{Overview of models and model comparison}
\label{sec:models}

As described in the Introduction, we consider three models with increasing level of complexity to fit the data. For each model, $\rho_i$ represents the energy density in the species $i$, $\rho_{\rm crit} = 3\MP^2H_0^2$, where $\MP$ is the reduced Planck mass, and we introduce the density parameter $\Omega_i = \rho_i/\rho_{\rm crit}$. The present energy density in radiation $\Omega_r = 5\times 10^{-5}$ is obtained from the CMB temperature measured from the 4-Year COBE-DMR CMB mission~\cite{Fixsen:1996nj} as $T_0 = \(2.728 \pm 0.004\)\,$K at 95\% C.L.. The fractional energy density in neutrinos $\Omega_\nu(z)$ assumes three active neutrinos of which two are massless and the third has a mass $m_{\nu_3} = 0.06\,$eV, as done in the baseline \textit{Planck} analyses.~\footnote{The value $0.06\,{\rm eV}$ is the minimal neutrino mass sum allowed within the normal ordering~\cite{deSalas:2017kay}, which is mildly favoured by current data~\cite{Vagnozzi:2017ovm,Simpson:2017qvj,Schwetz:2017fey,deSalas:2018bym}. Leaving the neutrino mass as a free parameter would not change our analysis substantially, given the current very tight upper limits on the sum of the neutrino masses~\cite{Palanque-Delabrouille:2015pga,Visinelli:2015uva,Cuesta:2015iho,Huang:2015wrx,Giusarma:2016phn,
Vagnozzi:2017ovm,Giusarma:2018jei,Aghanim:2018eyx,Vagnozzi:2019utt}.} We refer to the combination of invisible components as $\Omega_{\rm inv}(z) \equiv \Omega_r(1+z)^4 + \Omega_\nu(z)$. We treat baryons and CDM equivalently, both contributing to a total energy share $\Omega_M$. The remaining parameters are model-dependent and are discussed in the following.

\begin{enumerate}

\item $\Lambda$CDM model. Here, the normalized expansion rate is modelled as follows:
\be
	E_{{\Lambda}{\rm CDM}}(z) \!=\! \[\Omega_\Lambda + \Omega_M(1+z)^3 + \Omega_{\rm inv}(z)\]^{1/2}\,.
	\label{eq:ELCDM}
\ee
We fix the value of $\Omega_\Lambda$ by demanding that $E_{{\Lambda}{\rm CDM}}(0) = 1$, so that the parameter space associated to this model $\mathcal{S}_{{\Lambda}{\rm CDM}} = \{ \Omega_M, H_0, r_{\rm drag}, \ell_{\rm SN} \}$ is 4-dimensional.

\item $w$CDM model. Here, the normalized expansion rate is modelled as follows:
\be
	E_{w {\rm CDM}}(z) \!=\! \[\Omega_\phi(1 \!+\! z)^{3(1+w_\phi)} \!+\! \Omega_M(1 \!+\! z)^3 \!+\! \Omega_{\rm inv}(z)\]^{1/2}\,,
	\label{eq:wCDM}
\ee
where $w_{\phi}$ is a constant. We fix the value of $\Omega_\phi$ is fixed by demanding that $E_{w {\rm CDM}}(0) = 1$, so that the parameter space associated to this model $\mathcal{S}_{w {\rm CDM}} = \{ \Omega_M, w_\phi, H_0, r_{\rm drag}, \ell_{\rm SN} \}$ is 5-dimensional. Note that we allow $w_{\phi}$ to enter the phantom regime, where $w_{\phi}<-1$, since we are not restricting our attention to the case where the dark energy component is decribed by a single minimally coupled quintessence field with a canonical kinetic term.

\item $c$CDM model. Here, the normalized expansion rate is modelled as follows:
\bea
	E_{c{\rm CDM}}(z) = \bigg[\Omega_{\rm cc} + \Omega_\phi(1 + z)^{3(1+w_\phi)} + \Omega_M(1 + z)^3 + \Omega_{\rm inv}(z)\bigg]^{1/2}\,.
	\label{eq:cCDM}
\eea
We fix the value of $\Omega_\phi$ by demanding that $E_{w {\rm CDM}}(0) = 1$, so that the parameter space associated to this model $\mathcal{S}_{c{\rm CDM}} = \{ \Omega_{\rm cc}, \Omega_M, w_\phi, H_0, r_{\rm drag}, \ell_{\rm SN} \}$ is 6-dimensional. We demand that $\Omega_{\rm cc}$ be strictly negative by exploring the region of the parameter space $\Omega_{\rm cc} < 0$ when fitting the parameters in $\mathcal{S}_{c{\rm CDM}}$ against the data described in Sec.~\ref{sec:datasets}.

\end{enumerate}

To sample the posterior distribution of the parameters, we perform a Markov Chain Montecarlo (MCMC) analysis using the open-source Python package \texttt{emcee}~\cite{ForemanMackey:2012ig}. For each model $\mathcal{M}$ described in Sec.~\ref{sec:models}, we explore the associated parameter space $\mathcal{S}_{\mathcal{M}}$ by performing the analysis with a Metropolis-Hasting algorithm and assuming flat priors for all variables, except for the anchor for which we assume a Gaussian prior with mean and standard deviation given by the corresponding measurement. In more detail, we assume a flat prior for $\Omega_M \in \[0, 1\]$ and $\ell_{\rm SN} \in \[0.9, 1.2\]\,$Mpc for all models discussed, in addition to the DE EoS $w_\phi \in \[-2, -0.5\]$ for the $w$CDM and $c$CDM models, and the cosmological constant $\Omega_{\rm cc} \in \[-30, 0\]$ for the $c$CDM model. When anchoring the distance ladder to the comoving sound horizon at the drag epoch (Scenario I), we choose the flat prior $H_0 \in \[60, 80\]\,$km/s/Mpc and a Gaussian prior for $r_{\rm drag}$ based on the measurement by the \Planck collaboration. When the present Hubble rate is used as the anchor (Scenario II), we choose the flat prior $r_{\rm drag} \in \[120, 160\]\,$Mpc and a Gaussian prior for $H_0$ given by the SH0eS measurements~\cite{Riess:2019cxk}. We use the expression for the likelihood:
\be
	\mathcal{L} = \mathcal{L}_{\rm BAO} + \mathcal{L}_{\rm SN} + \mathcal{L}_{\rm Anchor}.
\ee

It is worth noting that the $\Lambda$CDM and $w$CDM models are \textit{nested} models, \textit{i.e.} the former is a particular case of the latter, recovered when setting $w=-1$. As such, we expect that the fit to data (quantified e.g. through the $\chi^2$) should not worsen when moving from the $\Lambda$CDM to the $w$CDM model. However, the same is not true for the $c$CDM model. In fact, within the $c$CDM model we require $\Omega_{cc}<0$, \textit{i.e.} that the cosmological constant be strictly negative. Therefore, it is not possible to recover neither the $\Lambda$CDM nor the $w$CDM model as a particular limit of $c$CDM. As a result, the fit to data will not necessarily improve when considering a negative cosmological constant.

At any rate, given that the $w$CDM and $c$CDM models possess respectively 1 and 2 extra parameters compared to the baseline $\Lambda$CDM model, it is important to assess whether the increased model complexity is warranted by a substantially better fit (if any) to the data. We perform a simple model comparison adopting the Akaike information criterion (AIC) to compare the competing models. The AIC of a given model is defined as~\cite{akaike}:
\begin{eqnarray}
{\rm AIC} = 2k+\min(\chi^2)\,,
\label{eq:aic}
\end{eqnarray}
where $k$ is the number of parameters of the model, and $\min(\chi^2)$ is the $\chi^2$ calculated at the best-fit point of the model. Therefore, $k=4\,,5\,,6$ for the $\Lambda$CDM, $w$CDM, and $c$CDM models respectively.

Assume we have two competing models $i$ and $j$, with AIC values ${\rm AIC}_i$ and ${\rm AIC}_j$ respectively, and assume that ${\rm AIC}_i<{\rm AIC}_j$. Then, model $i$ is to be preferred since it has a lower AIC. From Eq.~(\ref{eq:aic}), we see that one needs an improvement in fit of at least $\Delta \chi^2 \geq 2\Delta k$ in order to justify the increased number of parameters (for instance, $\Delta \chi^2=2$ when moving from $\Lambda$CDM to $w$CDM, or $\Delta \chi^2=4$ when moving to $c$CDM). Moreover, the quantity $\exp{[({\rm AIC}_i-{\rm AIC}_j)/2]}$ gives the relative likelihood of model $i$ with respect to model $j$.

\section{Results}
\label{sec:results}

We now discuss the results obtained analysing the datasets described in Sec.~\ref{sec:datasets} within the context of the three models described in Sec.~\ref{sec:models}, which we then compare using the AIC described in the same Section. In Tab.~\ref{fig:triangleplottable}, we show corner plots visualizing the 1D marginalized and 2D joint posterior distributions for the parameters of the models considered: $\Lambda$CDM model (upper panels), $w$CDM (middle panels), and $c$CDM model (lower panels), anchoring the BAO measurements either to $r_{\rm drag}$ as measured by \textit{Planck} (left column) or to $H_0$ as measured by SH0ES (right column). Constraints on the parameters of the three models are also presented in Tab.~\ref{table:results}.
\begin{table*}[!h]
	\sffamily
	\begin{center}
	\begin{tabular}{l*2{C}@{}}
	\toprule
	 & $r_{\rm drag}$ as anchor (Scenario I) & $H_0$ as anchor (Scenario II) \\ 
	\midrule
	 & \picLCDMrd & \picLCDMH  \\ 
	 & \picwCDMrd & \picwCDMH  \\ 
	 & \piccCDMrd & \piccCDMH \\ 
	\bottomrule 
	\end{tabular}
	\caption{Left column: corner plots showing the 1D marginalized and 2D joint posterior distributions for the parameters of the $\Lambda$CDM (top panel), $w$CDM (middle panel), and $c$CDM models (bottom panel), when the anchoring BAO measurements to $r_{\rm drag}$ as measured \textit{Planck}~\cite{Aghanim:2018eyx}. Right column: same as the left column, but anchoring BAO measurements to $H_0$ as measured by the SH0ES program~\cite{Riess:2019cxk}.}
	\label{fig:triangleplottable}
	\end{center}
\end{table*} 
\begin{table*}
\begin{center}
	{\renewcommand{\arraystretch}{1.4}
	\begin{tabular}{|l|l|l|l|l|l|l|l|l|}
	\hline
	Anchor & Model & $\Omega_c$ & $w_\phi$ & $\Omega_M$ & $H_0$ & $r_s$ & $\ell_{\rm SN}$ & $\Delta {\rm AIC}$ \\
	& & & & & km/s/Mpc & [Mpc] & [Mpc] & \\
	\hline
	$r_{\rm drag}$ & $\Lambda$CDM & & & $0.31 \pm 0.02$ & $68.53^{+0.79}_{-0.78}$ & $147.05^{+0.22}_{-0.21}$ & $1.04\pm 0.01$ & 0 \\ 
	$r_{\rm drag}$ & $w$CDM &  & $ -1.17^{+0.17}_{-0.19}$ & $0.36^{+0.04}_{-0.05}$ & $68.82^{+0.86}_{-0.86}$ & $147.06^{+0.21}_{-0.21}$ & $1.04\pm 0.01$ & 1.9 \\
	$r_{\rm drag}$ & $c$CDM & $> -13.88$ & $-1.02^{+0.02}_{-0.09}$ & $0.36^{+0.05}_{-0.05}$ & $68.83^{+0.86}_{-0.85}$ & $147.05^{+0.30}_{-0.31}$ & $1.04\pm 0.01$ & 3.6 \\
	\hline
	$H_0$ & $\Lambda$CDM & & & $0.31 \pm 0.02$ & $74.03^{+1.03}_{-1.01}$ & $136.16^{+2.43}_{-2.42}$ & $0.97\pm 0.01$ & 0 \\ 
	$H_0$ & $w$CDM &  & $ -1.18^{+0.17}_{-0.19}$ & $0.36^{+0.04}_{-0.05}$ & $74.04^{+0.99}_{-1.01}$ & $136.72^{+2.52}_{-2.43}$ & $0.97\pm 0.01$ & 1.3 \\ 
	$H_0$ & $c$CDM & $> -14.48$ & $-1.02^{+0.02}_{-0.09}$ & $0.36^{+0.05}_{-0.05}$ & $73.96^{+1.41}_{-1.45}$ & $136.86^{+3.22}_{-3.04}$ & $0.97\pm 0.02$ & 4.8 \\
	\hline
	\end{tabular}}
	\caption{Constraints on the parameters of the three models considered in this work, obtained using the two different choices of anchor (specified in the first column). We report 68\%~C.L. intervals on all parameters, except for $\Omega_{c}$ for which we do not have a detection and hence report the  95\%~C.L. lower bound. The last column reports $\Delta {\rm AIC}$, defined in Eq.~(\ref{eq:aic}), and calculated with respect to the $\Lambda$CDM model. A positive value of $\Delta {\rm AIC}$ indicates a preference for $\Lambda$CDM.}
	\label{table:results}
	\normalsize
\end{center}
\end{table*}

As we see from Tab.~\ref{table:results}, our analysis does not reveal any evidence for a non-zero negative cosmological constant. In fact, when analysing our datasets within the $c$CDM model, we only obtain a lower limit of $\Omega_{cc} \gtrsim -14$ regardless of whether we use $r_{\rm drag}$ as measured by \textit{Planck} or $H_0$ as measured by SH0ES as anchor. Moreover, within both the $w$CDM and $c$CDM models, we see a $1\sigma$ preference for a phantom DE component ($w_{\phi}<-1$), indicating that data prefers a late-time expansion rate lower than the one obtained when DE is entirely in the form of a (positive) cosmological constant. We also see that when moving from the $w$CDM model to the $c$CDM model the central value of $w_{\phi}$, although still phantom, moves upwards towards $w=-1$. The reason is that, as discussed in Sec.~\ref{sec:intro}, introducing a negative cosmological constant lowers the expansion rate even more drastically than when considering a phantom DE component, lessening the need for the latter and explaining why $w_{\phi}$ moves towards $-1$.

It is also worth noting that within the $w$CDM and $c$CDM model we infer a higher matter density than within $\Lambda$CDM (although with uncertainty about twice as large). The reason is again that a phantom dark energy component (with the addition of a negative cosmological constant) lowers the expansion rate at late times, which can be compensated by increasing the matter density. This result is analogous to the one found in~\cite{Vagnozzi:2018jhn}, where it was shown that in models with $w>-1$ one infers a lower matter density, because the late-time expansion rate in this case is higher, leaving less room for matter components.

We also notice that when anchoring BAO measurements to $r_{\rm drag}$ as measured by \textit{Planck}, we recover a value of $H_0$ which is essentially $H_0 \approx 68.5\,{\rm km}\,{\rm s}^{-1}\,{\rm Mpc}^{-1}$ for all three models, which indicates that the $c$CDM model has not been able to alleviate the $H_0$ tension. Finally, we note that anchoring BAO measurements to $H_0$ as measured by SH0ES, we infer a value for $r_{\rm drag}$ which is about $8\%$ lower than that obtained when anchoring to $r_{\rm drag}$ as measured by \textit{Planck}, in full agreement with earlier results~\cite{Bernal:2016gxb,Lemos:2018smw,Aylor:2018drw}.

One further interesting point worth noting from Tab.~\ref{table:results} is that the values we infer for certain parameters (such as $\Omega_M$, $H_0$, $r_s$, and $l_{\rm SN}$) are essentially the same across different models (particularly for the $w$CDM and $c$CDM models) and for different choices of anchors. One particularly striking case, for example, is $\Omega_M$, for which we basically infer $\Omega_M=0.36 \pm 0.05$ for both the $w$CDM and $c$CDM models, regardless of the choice of anchor. On the other hand, for the $\Lambda$CDM model we inferred $\Omega_M=0.31 \pm 0.02$. This is consistent with the fact that we do not have a detection of non-zero negative cosmological constant, and therefore from the point of view of the data we used there is not much different between the $w$CDM and $c$CDM models. On the other hand, parameters such as $\Omega_M$ are extremely well determined by BAO data, independently of the choice of anchor. Similar considerations hold for the other parameters whose values inferred within the $w$CDM and $c$CDM models are very similar to one another regardless of the choice of anchor.

We now compare the three models using the Akaike information criterion introduced in Sec.~\ref{sec:models}, beginning with the case when BAO measurements are anchored to $r_{\rm drag}$. In this case, we find minimal improvements in the best-fit $\chi^2$ when moving away from the $\Lambda$CDM model. In particular, we find $\Delta \chi^2=-0.1$ and $\Delta \chi^2=-0.4$ for the $w$CDM and $c$CDM models respectively. According to Eq.~(\ref{eq:aic}), these values correspond to $\Delta {\rm AIC}=1.9$ and $\Delta {\rm AIC}=3.6$ for the $w$CDM and $c$CDM models respectively. In both cases $\Delta {\rm AIC}>0$, indicating that the tiny improvement in fit does not warrant the addition of extra parameters (1 extra parameter and 2 extra parameters for $w$CDM and $c$CDM respectively), and therefore that the baseline $\Lambda$CDM model is preferred from a statistical point of view. In fact, we find that the relative likelihoods of the $w$CDM and $c$CDM models over $\Lambda$CDM are $\approx 0.38$ and $\approx 0.16$ respectively.

We find completely analogous, if not less optimistic results, when anchoring BAO measurements to $H_0$. In this case, we find $\Delta \chi^2=-0.7$ and $\Delta \chi^2=+0.8$ for the $w$CDM and $c$CDM models respectively. We see that within the $c$CDM model the quality of the fit has actually worsened. According to Eq.~(\ref{eq:aic}), these values correspond to $\Delta {\rm AIC}=1.3$ and $\Delta {\rm AIC}=4.8$ for the $w$CDM and $c$CDM models respectively. Again, in both cases $\Delta {\rm AIC}>0$, with relative likelihoods for the $w$CDM and $c$CDM models over the $\Lambda$CDM model being $\approx 0.52$ and $\approx 0.09$ respectively.

\section{Conclusions}
\label{sec:summary}

In this work we have revisited the possibility that the dark energy sector might feature two components: a negative cosmological constant, and a component with positive energy density (such as a quintessence field, but with the possibility that is equation of state might be phantom) on top. While such a model (which we referred to as $c$CDM model) is certainly a toy model, it can be seen as a proxy for more realistic string-inspired models~\cite{Vafa:2005ui,Danielsson:2018ztv,Obied:2018sgi,Ooguri:2018wrx,Palti:2019pca}, where it is rather natural to have an AdS background with on top a large number of light bosons corresponding to moduli determining the size and shape of the extra dimensions. We have tested our toy $c$CDM model against low-redshift distance measurements from a collection of recent BAO surveys and the Pantheon SNeIa catalogue. To interpret our BAO measurements, we require either an anchor either at high or low redshift. We have experimented with two different choices of anchor: a high-redshift anchor based on the sound horizon at the drag epoch $r_{\rm drag}$ as determined by the \textit{Planck} collaboration~\cite{Aghanim:2018eyx}, or a low-redshift anchor based on the local distance ladder measurement of the Hubble constant $H_0$ measured by the SH0ES program~\cite{Riess:2019cxk}.

Our results are summarized in Tab.~\ref{fig:triangleplottable} and Tab.~\ref{table:results}. We have found no evidence for a negative cosmological constant, but only obtained the very loose lower bound $\Omega_c \gtrsim -14$. Moreover, we find a mild preference for a phantom equation of state ($w_{\phi}<-1$) for the dark energy component on top of the negative cosmological constant, in agreement with recent works.

We have also compared the three models considered in this work: the baseline $\Lambda$CDM model, the one-parameter $w$CDM extension where we allow the DE EoS $w_{\phi}$ to vary freely, and finally the $c$CDM model with a negative cosmological constant in addition. We have found that for most choices of anchor/model the increased model complexity only marginally improves the quality of the fit (which actually deteriorates when considering the $c$CDM model and anchoring BAO measurements to $H_0$). When accounting for the increased model complexity by comparing the three models using the Akaike information criterion, we find that the baseline $\Lambda$CDM model is always statistically favoured over the extensions we considered.

The results of our work are less optimistic than earlier results in e.g. Refs.~\cite{Poulin:2018zxs,Wang:2018fng,Dutta:2018vmq}, which found indications for a negative dark energy density at $z \approx 2$. There are nonetheless some fundamental differences between the approaches of Refs.~\cite{Poulin:2018zxs,Wang:2018fng,Dutta:2018vmq} and ours. We have not attempted to non-parametrically reconstruct the dark energy density, but rather focused on a string-inspired toy model which we have then fitted to data. Moreover, on top of the negative cosmological constant we have considered a dark energy component with constant equation of state, although an obvious extension would be to consider a time-varying equation of state (which is very natural for a quintessence model). In any case, our results suggest that a negative cosmological constant is certainly consistent with data. Should the trend which sees low-redshift data favouring a lower expansion rate continue (for instance in light of the persisting $H_0$ tension), our scenario featuring a negative cosmological constant with a quintessence component on top might be worth reconsidering. It is worth reminding once more that, while being a toy model, our $c$CDM model nonetheless enjoys a fundamental string-inspired motivation.

Finally, there are several avenues by which we can extend and improve the current work. From the observational side, an analysis performing a full fit to CMB data from the \textit{Planck} satellite, following the upcoming public release of the 2019 legacy likelihood, would be valuable. It would also be worth allowing for more freedom in the dark energy sector, for instance by considering a time-varying dark energy component on top of the negative cosmological constant (in this work we have restricted ourselves to the case where the equation of state is a constant). Finally, it would also be interesting to consider forecasts for how our constraints could improve when data from future CMB experiments will become available (such as \textit{Simons Observatory}~\cite{Ade:2018sbj,Abitbol:2019nhf} or CMB-S4~\cite{Abazajian:2016yjj}), or gravitational wave standard sirens will improve constraints on the late-time expansion rate (e.g. along the lines of Ref.~\cite{Du:2018tia}). On the theory side, it would be worth further refining our toy model by more carefully modelling the negative cosmological constant and quintessence component emerging from realistic string constructions. We plan to report on these and other issues in upcoming work.

\begin{acknowledgments}
We thank Massimiliano Lattanzi for useful discussions and suggestions that improved the manuscript. L.V. and S.V. acknowledge support by the Vetenskapsr\r{a}det (Swedish Research Council) through contract No. 638-2013-8993 and the Oskar Klein Centre for Cosmoparticle Physics. S.V. acknowledges support from the Isaac Newton Trust and the Kavli Foundation through a Newton-Kavli fellowship. U.D. acknowledges support by the Vetenskapsr\r{a}det (Swedish Research Council) under contract 2015-04814. L.V. acknowledges the kind hospitality of the INFN Laboratori Nazionali di Frascati and the Leinweber Center for Theoretical Physics, where part of this work was carried out.
\end{acknowledgments}

\bibliography{NegativeCC.bib}

\end{document}